\documentclass[12pt,preprint]{aastex}

\usepackage{natbib,graphicx,epsfig}
\usepackage{color}

\newcommand{\psim}{\lower.5ex\hbox{$\; \buildrel \propto \over\sim \;$}}

\newcommand{\gr}{$\gamma$-ray }

\shorttitle{High-Energy Gamma-Ray Emission from Starburst Galaxies}
\shortauthors{Abdo et al.}

\begin{document}

\title{Detection of Gamma-Ray Emission from the Starburst Galaxies M82 and NGC 253 with the Large Area Telescope on 
{\it Fermi}}




\author{
A.~A.~Abdo\altaffilmark{2,3}, 
M.~Ackermann\altaffilmark{4}, 
M.~Ajello\altaffilmark{4}, 
W.~B.~Atwood\altaffilmark{5}, 
M.~Axelsson\altaffilmark{6,7}, 
L.~Baldini\altaffilmark{8}, 
J.~Ballet\altaffilmark{9}, 
G.~Barbiellini\altaffilmark{10,11}, 
D.~Bastieri\altaffilmark{12,13}, 
K.~Bechtol\altaffilmark{4}
% ...
}
% ...
\altaffiltext{2}{Space Science Division, Naval Research Laboratory, Washington, DC 20375, USA}
\altaffiltext{3}{National Research Council Research Associate, National Academy of Sciences, Washington, DC 20001, USA}
\altaffiltext{4}{W. W. Hansen Experimental Physics Laboratory, Kavli Institute for Particle Astrophysics and Cosmology, Department of Physics and SLAC National Accelerator Laboratory, Stanford University, Stanford, CA 94305, USA}
\altaffiltext{5}{Santa Cruz Institute for Particle Physics, Department of Physics and Department of Astronomy and Astrophysics, University of California at Santa Cruz, Santa Cruz, CA 95064, USA}
\altaffiltext{6}{Department of Astronomy, Stockholm University, SE-106 91 Stockholm, Sweden}
\altaffiltext{7}{The Oskar Klein Centre for Cosmo Particle Physics, AlbaNova, SE-106 91 Stockholm, Sweden}
\altaffiltext{8}{Istituto Nazionale di Fisica Nucleare, Sezione di Pisa, I-56127 Pisa, Italy}
\altaffiltext{9}{Laboratoire AIM, CEA-IRFU/CNRS/Universit\'e Paris Diderot, Service d'Astrophysique, CEA Saclay, 91191 Gif sur Yvette, France}
\altaffiltext{10}{Istituto Nazionale di Fisica Nucleare, Sezione di Trieste, I-34127 Trieste, Italy}
\altaffiltext{11}{Dipartimento di Fisica, Universit\`a di Trieste, I-34127 Trieste, Italy}
\altaffiltext{12}{Istituto Nazionale di Fisica Nucleare, Sezione di Padova, I-35131 Padova, Italy}
\altaffiltext{13}{Dipartimento di Fisica ``G. Galilei", Universit\`a di Padova, I-35131 Padova, Italy}
% ...


\begin{abstract}
We report the detection of high-energy \gr emission from two starburst galaxies using data obtained with the Large Area Telescope on board the {\it Fermi} Gamma-ray Space Telescope. Steady point-like emission above 200 MeV has been detected at significance levels of $6.8\sigma $  and $4.8\sigma$  respectively, from sources positionally coincident with locations of the starburst galaxies M82 and NGC 253. The total fluxes of the sources are consistent with \gr emission originating from the interaction of cosmic rays with local interstellar gas and radiation fields and constitute evidence for a link between massive star formation and \gr emission in star-forming galaxies. 
\end{abstract}




\section{Introduction}

Cosmic rays are believed to be accelerated by supernova remnant shocks
that are formed when a star explodes (Ginzburg \& Syravotskii 1964;
Hayakawa 1969). Observations of $\gamma$ rays from supernova remnants in
the Milky Way would apparently offer the best opportunity to identify
the sources of cosmic rays, but cosmic-ray diffusion throughout the
Galaxy results in a bright \gr glow, making it difficult to attribute
$\gamma$ rays to cosmic-ray electrons, protons or ions accelerated by Galactic supernova remnants. Direct evidence for the sources of cosmic rays is therefore still lacking. 

The supernova remnant paradigm for cosmic-ray origin can also be tested by measuring the \gr emission from star-forming galaxies. Starburst galaxies, in particular, should have larger \gr intensities compared to the Milky Way due to their increased star-formation rates and greater amounts of gas and dust that reprocess light into the IR, and, with photons, serve as targets for \gr production by cosmic ray electrons and ions. If the \gr production rate is sufficiently increased, star-forming galaxies will be detectable by the current generation of instruments, as early estimates (e.g., V\"olk et al.\ 1989, 1996; Aky\"uz et al.\ 1991; Paglione et al.\ 1996) and recent detailed models (e.g., Domingo-Santamar\'ia \& Torres 2005; Persic et al.\ 2008; de Cea del Pozo et al.\ 2009; Rephaeli et al.\ 2009; Lacki et al.\ 2009) predict. 

Here we report the detection of the starburst galaxies M82 and NGC 253
in high-energy $\gamma$ rays from observations with the Large Area
Telescope (LAT) on board the {\it Fermi} Gamma-ray Space Telescope. A description of the analysis of the observations is given in Section 2. In Section 3, the measured spectra and fluxes are compared with predictions based on theories of cosmic-ray origin from supernovae in star-forming galaxies. 

\section{Observations and Analysis}

The LAT is a pair-conversion telescope with a precision tracker and
calorimeter, a segmented anti-coincidence detector (ACD) which covers
the tracker array, and a programmable trigger and data acquisition
system. Incoming $\gamma$ rays convert into electron-positron pairs while
traversing the LAT. The directions of primary $\gamma$ rays are
reconstructed using information provided by the tracker subsystem
while the energies are measured via the calorimeter subsystem. The ACD subsystem vetoes the great majority of
cosmic rays that trigger the LAT. The energy range of the LAT spans from 20
MeV to $> 300$ GeV with an angular resolution of approximately
$5.1^\circ$ at 100 MeV and narrowing to about $0.14^\circ$ at 10
GeV\footnote{Angular resolution is defined here as the 68\% containment radius of the LAT point spread
function averaged over the intrument acceptance and including photons
which convert in either the thick or thin layers of the tracker
array.}. Full details of the instrument, onboard and ground data processing, and other mission-oriented support are given in Atwood et al.\ (2009).

The LAT normally operates in a scanning mode (the `sky survey' mode)
that covers the whole sky every two orbits (i.e., $\sim 3$ hrs). We
use data taken in this mode from the commencement of scientific
operations in early-August 2008 to early-July 2009. The data were
prepared using the LAT Science Tools package. Only events satisfying
the standard low-background event selection (the so-called `Diffuse'
class events corresponding to the P6V3 instrument response functions
described in Rando (2009)) and coming from zenith angles $< 105^\circ$ (to greatly reduce the contribution by Earth albedo $\gamma$ rays) were used in the present analysis. To further reduce the effect of Earth albedo backgrounds, time intervals when the Earth was appreciably in the field of view (specifically, when the center of the field of view was more than $43^\circ$ from the zenith) were also excluded from the analysis.

\begin{figure}[t]
\center
\includegraphics[width = 7.0cm]{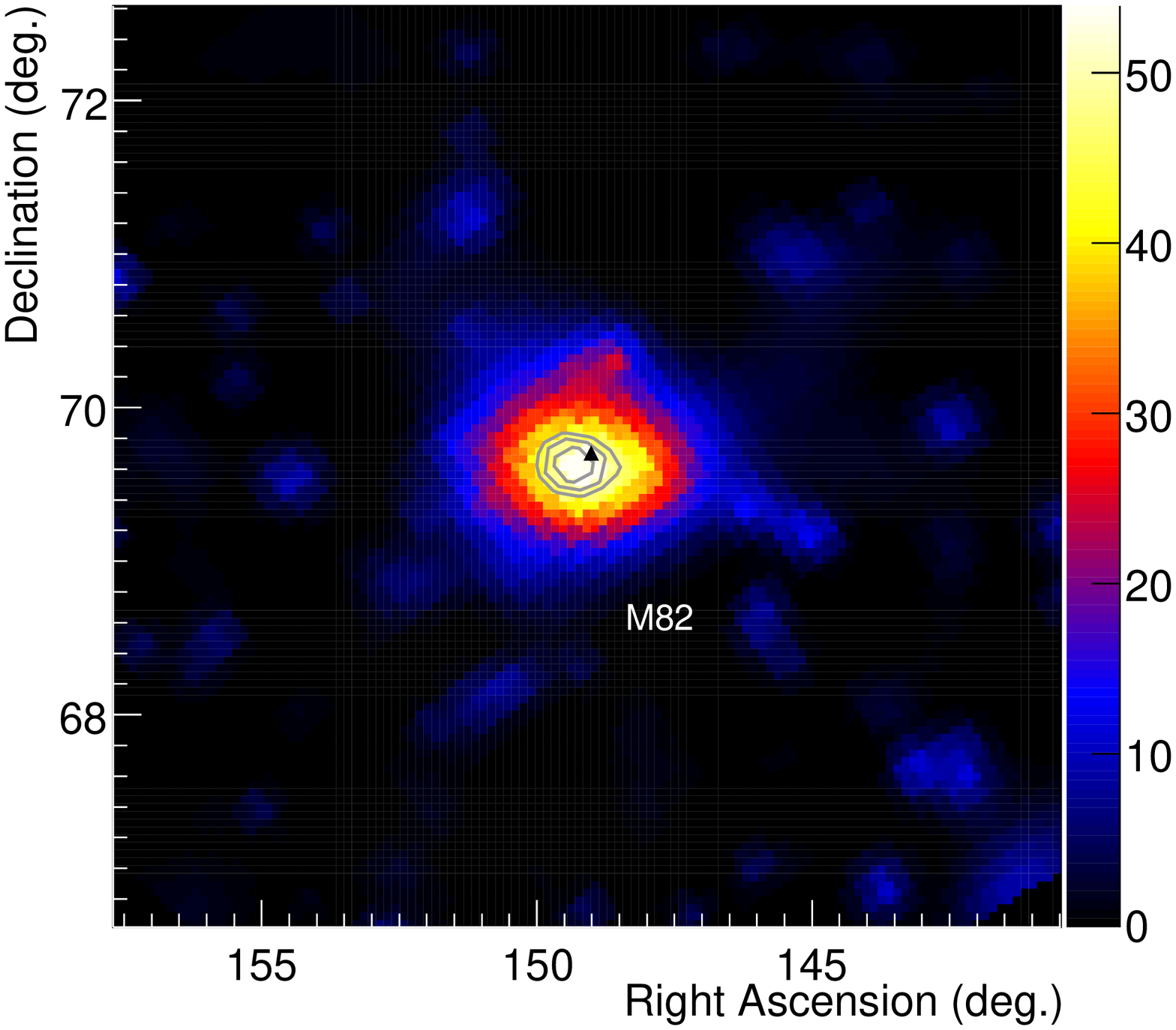}
\includegraphics[width = 7.0cm]{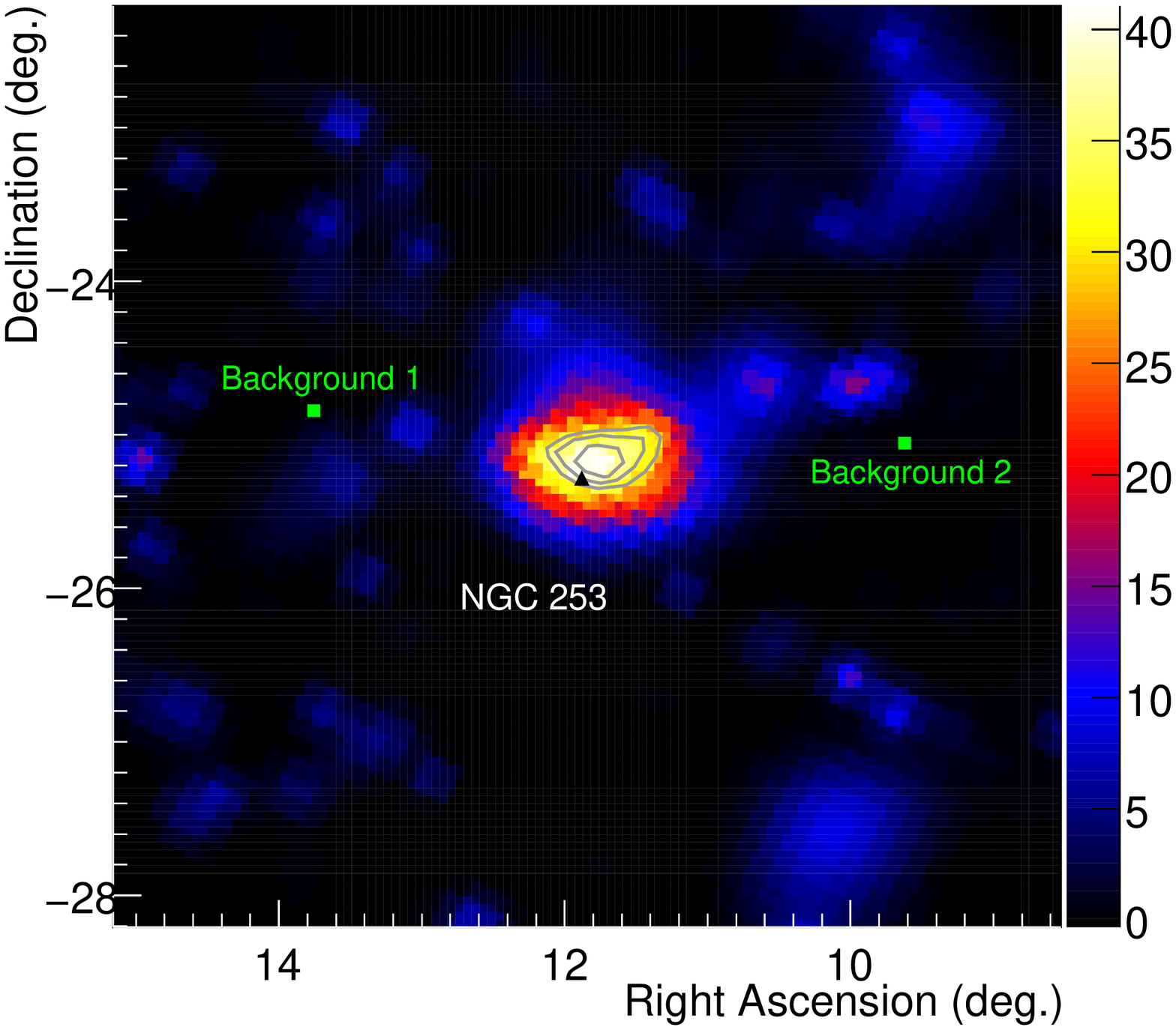}
\caption{Test statistic maps obtained from photons above 200 MeV
  showing the celestial regions ($6^\circ$ by $6^\circ$) around M82 and NGC
  253. Aside from the source associated with each galaxy, all other
  {\it Fermi}-detected sources within a $10^\circ$ radius of the
  best-fit position have been included in the background model as well
  as components describing the diffuse Galactic and isotropic \gr emissions. Black triangles denote
  the positions of M82 and NGC 253 at optical wavelengths; gray lines indicate the 0.68, 0.95, and 0.99 confidence level contours on the position of the observed \gr excess; green squares show the positions of individual background sources. The color scale indicates the point-source test statistic value at each location on the sky, proportional to the logarithm of the likelihood ratio between a \gr point-source hypothesis ($L_1$) versus the null hypothesis of pure background ($L_0$); $TS\equiv 2 (\ln L_1- \ln L_0 )$ (Mattox et al. 1996). }
\vskip0.2in
\label{tsmap}
\end{figure}

We use all $\gamma$ rays with energy $> 200$ MeV within a $10^\circ$
radius region of interest (ROI) of the optical locations for the
galaxies M82 and NGC 253. Detection significance maps for each ROI are
shown in Fig. \ref{tsmap}. The background model for each ROI includes
all LAT-detected sources along with components describing the diffuse Galactic and isotropic \gr emissions. Each map shows a bright and isolated \gr excess above the background that is consistent with the location of the nominal (optical) position of the respective starburst galaxy. 
  
The data were analyzed using the LAT Science Tools package (v9r15p2),
which is available from the {\it Fermi} Science Support Center, using
P6V3 post-launch instrument response functions (IRFs). These IRFs take
into account event pile-up and accidental coincidence effects in the
detector subsystems that were not considered in the definition of the
pre-launch IRFs. We used a maximum likelihood fitting procedure ({\it gtlike}) to determine the positions of the \gr sources associated with M82 and NGC 253 (see Table~\ref{fit-results}). The angular separation between the best-fit location and the core of each galaxy is $0.05^\circ$ for M82 and $0.12^\circ$ for NGC 253. Systematic uncertainties in the positions due to inaccuracies in the point spread function and telescope alignment are estimated to be less than $0.01^\circ$.



We tested the possibility that the sources are spatially extended by fitting
two-dimensional Gaussian-shaped intensity profiles. The widths and
locations of the profiles were adjusted and refit over the region in
an iterative procedure but we found no significant evidence for source
extension in our data. We verified these results using a
likelihood fitting procedure capable of modeling spatially extended
\gr sources ({\it sourcelike}). A comparison between the point
and extended source hypotheses using this method produces negligible
changes in detection significance. From our analysis we set
upper limits on the angular sizes of the emitting regions as
$0.18^\circ$ for M82 and $0.30^\circ$ for NGC 253 at the 95\%
confidence level assuming a two-dimensional Guassian spatial model
parameterized by the 68\% surface intensity containment radius. By
comparison, the angular sizes of the galaxies are $0.19^\circ\times
0.07^\circ$ for M82 and $0.46^\circ\times
0.11^\circ$ for NGC 253 as measured in the ultraviolet band (Gil de
Paz et al. 2007). The starburst cores of M82 (V\"olk et al. 1996) and
NGC 253 (Ulvestad 2000) have an angular extent $<0.01^\circ$ and cannot be resolved by the
LAT.

Spectral analysis is a separate maximum likelihood calculation for
which we have adopted the point-source hypothesis and best-fit position determined during the
localization and extension fitting step.


Diffuse \gr emission from the Milky Way is treated with the
Galactic diffuse model described within the gll\_iem\_v02.fit file
suitable for analysis with the Science Tool {\it gtlike}. In addition to the
spatially structured Galactic diffuse emission, {\it Fermi} also observes an
isotropic diffuse component which includes both extragalactic diffuse
\gr emission and instrumental background from charged particles triggering the LAT. The isotropic diffuse emission has been treated with
isotropic\_iem\_v02.txt. Excepting the sources associated with M82
and NGC 253, all individual objects detected by {\it Fermi} after 11
months of scientific operations within a $10^\circ$ radius of
the best-fit position of each galaxy are also included into the
background description of each region as distinct point sources.


We considered alternative associations for the two LAT sources of
interest aside from M82 and NGC 253 in the CRATES catalog of flat-spectrum radio
sources (14467 entries, Healey et al.\ 2007) and the Candidate
Gamma-Ray Blazar Survey catalog, CGRaBS (1625 entries, Healey et al.\
2008). Both of these catalogs show high correlation with
\gr bright blazars based on multiwavelength observations. However,
there are no likely CRATES or CGRaBs objects within the positional uncertainty of
either LAT source. Near NGC 253, the only
source of possible concern is a $\sim 40$ mJy NVSS (Condon et
al. 1998) radio source at 1.4 GHz with unknown spectrum. Such a source would be unusually weak by comparison with the radio fluxes of LAT blazars. 


We searched for flux variability for each \gr source by creating
monthly flux histories of the total photon flux $> 400$ MeV arriving from within a
circular regions $1^\circ$ in radius centered on the {\it Fermi}-determined
locations. No flaring events are observed and the $\chi^2$
goodness-of-fit test is consistent with constant flux for each source
(reduced $\chi^2$ = 0.80 and 1.03 for M82 and NGC 253, respectively,
each with 9 degrees of freedom). Lack of variability is in accord with the cosmic-ray origin
hypothesis where most of the emission derives from diffuse cosmic-ray
interactions, though mild variability of $\gamma$ rays and radio
emission (Kronberg et al.\ 2000, Brunthaler et al. 2009b) might still occur if M82 or NGC 253 had
a recent supernova. Large amplitude \gr variability on short
timescales would rule out a cosmic-ray origin of the $\gamma$ radiation.

Table~\ref{fit-results} summarizes the results of the analyses of M82
and NGC 253. The overall detection significance is $6.8\sigma$ for
M82 and $4.8\sigma$  for NGC 253. Note that the significance level for
these moderately hard spectrum sources is based on the number of high
energy photons compared to the expected background, whereas the flux
uncertainty is based on the number of such photons, which is not
large, and systematic effects. The integral photon fluxes over 100 MeV
are calculated by extrapolation of the fitted spectral models.

\section{Interpretation}

With the nearest luminous starburst galaxies, M82 and NGC 253,
detected by the {\it Fermi} Gamma-ray Space Telescope, we can test
long-standing predictions based on the cosmic-ray paradigm that
diffuse \gr emission from star-forming galaxies is produced via
cosmic-ray interactions. The distance to M82 is $3.63 \pm 0.34$ Mpc
(Freedman  et al. 1994), and distance estimates to NGC 253 range from 2.5 Mpc (Turner \& Ho 1985, Mauersberger et
al. 1996) to $3.9 \pm 0.37$ Mpc (Karachentsev et al. 2003). Vigorous
star formation is observed within the central several hundred parsecs of these galaxies. Estimates of the supernova
explosion (SN) rate vary from $\approx 0.08$ -- $0.3$ yr$^{-1}$ in
M82, to $\approx 0.1$ -- $0.3$ yr$^{-1}$ in NGC 253, 
compared to the supernova rate of $\approx 0.02$ yr$^{-1}$ in the Milky
Way. Recent studies of  M82 find $7\times 10^8 M_\odot$ in atomic H~I
gas and $1.8 \times 10^9 M_\odot$ in H$_2$ gas (Casasola et
al. 2004). The central region of NGC 253 contains a bar of molecular gas with an estimated mass of $4.8\times
10^8 M_\odot$ (Canzian et al. 1988), and its total gas content is
$\approx 60$\% of the Milky Way's (Boomsma et al. 2005; Houghton et
al. 1997; Brunthaler et al. 2009b), reflecting active star formation
taking place in these relatively small galaxies. 

Table~\ref{gal-comp} gives adopted values of distance $d$, supernova rate $R_{SN}$,
total gas mass $M_{Gas}$, \gr flux $F(>100$ MeV), and \gr
luminosities for M82 and NGC 253, alongside those of the Large
Magellanic Cloud (LMC) and the Milky
Way. The 100 MeV to 5 GeV \gr luminosity of M82 and NGC 253 is $\approx 10^{40}$ erg s$^{-1}$, 
compared to $\approx 3\times 10^{39}$ erg s$^{-1}$ for the Milky Way,
and $\approx 4.1\times 10^{38}$ erg s$^{-1}$ for the LMC. These
galaxies lack active central nuclei and so require a different origin
for their \gr fluxes than from galaxies with supermassive
black-hole jets. The $\gamma$ rays from our Galaxy and the LMC
arise predominantly from cosmic rays interacting with
interstellar gas and radiation fields. The starburst galaxies M82 and
NGC 253, though having less gas than the Milky Way, have a factor 2 -- 4 greater \gr luminosity,
suggesting a connection between active star formation and enhanced cosmic-ray energy densities in star-forming galaxies. 

\begin{figure}[t]
\center
\includegraphics[width= 16.0cm]{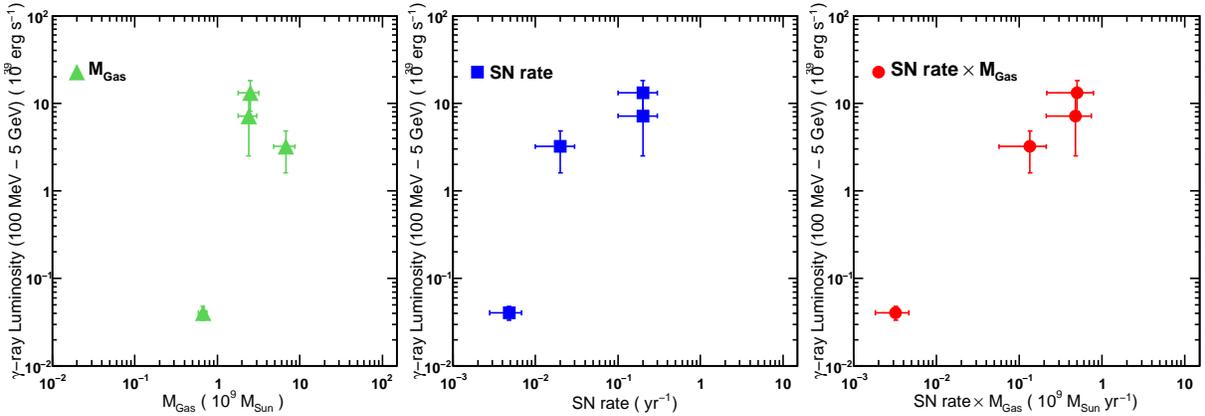}
\caption{Relationship between supernova rate, total gas mass, and
  total \gr luminosity of four galaxies detected by their
  diffuse high-enery emission. In order of ascending
  \gr luminosity, the plotted galaxies are the LMC,
  Milky Way, NGC 253, and M82. Three panels are shown to compare
  different possible correlations with the \gr luminosity: total
  gas mass (left), supernova rate (center), and product of the total
  gas mass and supernova rate (right). This figure is based upon the observed
  quantities and associated uncertainties presented in
  Table~\ref{gal-comp}.}
\vskip0.2in
\label{comparison-plot}
\end{figure}

We examine several possible
correlations between total gas mass, supernova rate, and \gr
luminosity of these four galaxies as illustrated in Fig. \ref{comparison-plot} (cf. Pavlidou \& Fields 2001, for local group galaxies).
In the left-hand panel, we find a poor correlation between \gr
luminosity and gas mass, and a weak linear correlation between \gr
luminosity and supernova rate. Models that attribute the $\gamma$ rays to
cosmic-ray processes depend both on enhanced cosmic-ray intensities, which 
depends on the supernova rate, and
large quantities of target gas, suggesting that
the \gr luminosity is proportional to the product of the total supernova
rate and gas mass, as shown in the right-hand panel of Fig. \ref{comparison-plot}. 
Note that while the detection of galaxies in this sample is flux-limited,
the measured gas masses and supernova rates for all galaxies are not,
so that the dependence of \gr luminosity on these parameters
reflect underlying physical relationships rather than sensitivity effects.
Although the sample size is small, this result argues in favor of a scaling
of \gr luminosity according to expectations from the hypothesis that the
emission is produced by cosmic-ray interactions.  

Evaluation of the dependence of \gr luminosity on galaxy properties is complicated, however, by 
star formation rates that depend on location in the galaxy. 
Radio and infrared observations reveal that the starburst activity in M82 and NGC 253 takes place in a
relatively small central region, radius $\sim$200 pc for both M82 (V\"olk et
al. 1996) and NGC 253 (Ulvestad 2000), so that the distribution of the
cosmic rays in the galaxies is probably not uniform. 
In cases where \gr emission can be resolved, as for the Milky Way,
this  can be seen directly  (Dragicevich et al.\ 1999). For instance, \gr emission
from the LMC is mostly produced in the star-forming
region 30 Doradus, and does not simply trace star formation and total gas
mass (Abdo et al. 2009a). 

\begin{figure}[t]
\center
\includegraphics[width= 8.0cm]{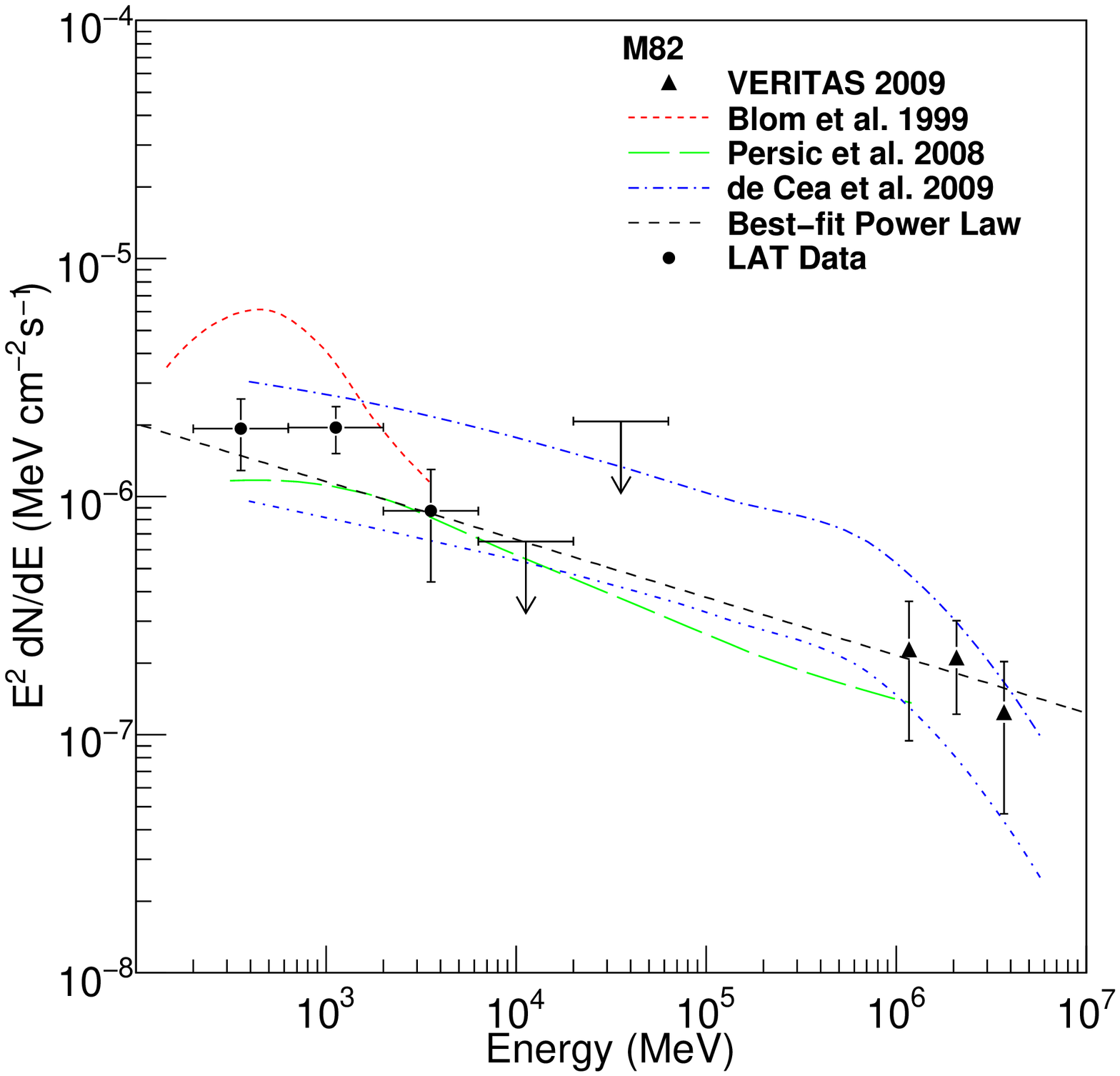}
\includegraphics[width= 8.0cm]{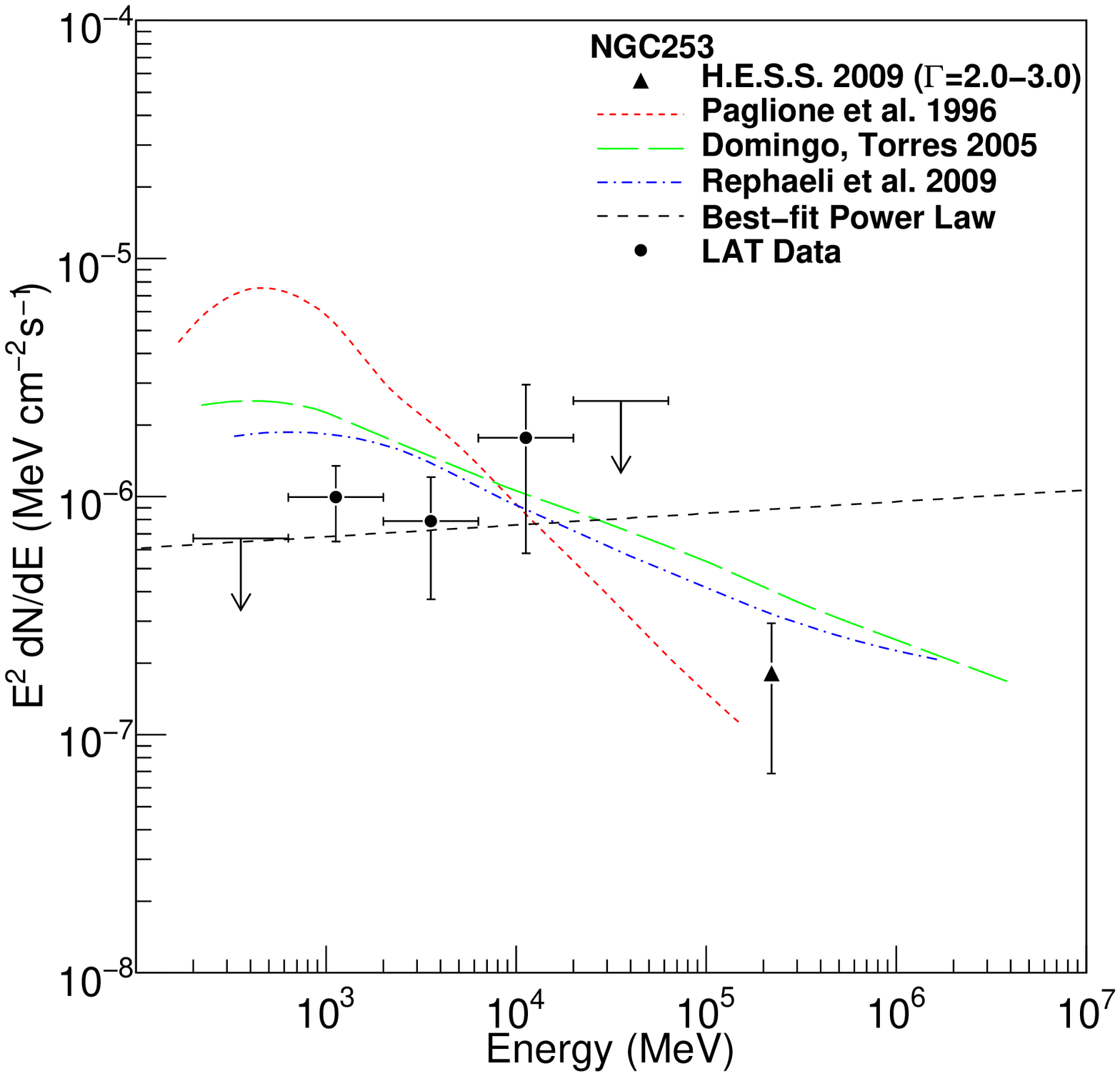}
\caption{Spectral energy distributions of M82 and NGC 253. The spectra
  were obtained using {\it gtlike}, with flux points extracted based
  upon the parameters presented in Table~\ref{fit-results}. Upper limits from the LAT correspond to the 0.68 confidence
  level. Three flux points in the TeV energy range are provided by VERITAS
  observations of M82 (Acciari et al. 2009). The single very
  high energy flux point for NGC 253 is computed from the integral
  photon flux over 220 GeV reported by the H.E.S.S. collaboration (Acero et
  al. 2009) and assumes a power-law spectral model marginalized over
  photon indices ranging from 2.0 to 3.0. Several theoretical predictions are plotted
  for comparison to the observed \gr spectra.}
\vskip0.2in
\label{SED}
\end{figure}

Theoretical predictions, despite using different assumptions and
treating the processes with varying levels of detail, are largely
consistent with the detected integral flux of M82 (e.g. V\"olk et
al. 1989; Ak\"uz et al. 1991; Persic et al. 2008; de Cea et al. 2009)
and NGC 253 (e.g. Paglione et al. 1996; Domingo-Santamar\'ia \& Torres
2005; Persic et al. 2008). Fig. \ref{SED} shows the predicted and
observed spectra. In the case of NGC 253, the predicted photon flux
$(>100$ MeV) is $2.3\times 10^{-8}$ photons cm$^{-2}$ s$^{-1}$
(Domingo-Santamar\'ia \& Torres 2005) and $2\times 10^{-8}$ photons
cm$^{-2}$ s$^{-1}$ (Persic et al. 2008). For M82, the predicted photon
flux ($>100$ MeV) is between $2.6 \times 10^{-8}$ and $8.3 \times
10^{-9}$ photons cm$^{-2}$ s$^{-1}$ (de Cea et al. 2009) due to
systematic uncertainties in model parameters; and $\approx 10^{-8}$
photons cm$^{-2}$ s$^{-1}$ (Persic et al. 2008). Furthermore, extrapolation of the
best-fit power-law spectral model at GeV energies provides a smooth connection to
flux densities of M82 reported at TeV energies (Acciari et
al. 2009). Although not highly constraining due to the faintness of
M82 in the GeV band, the fitted spectrum suggests that a single physical emission mechanism
dominates from GeV to TeV energies. The relationship between the GeV and
TeV emission for NGC 253 is less clear given the current data. Also,
note that the inner starburst region of NGC 253 has about
a factor of 3 less radio flux than that of M82 at 1.4 GHz, consistent with the
galaxy being less luminous in $\gamma$ rays (M82, Klein et al. 1988; NGC
253, Carilli 1996).

The star-forming galaxy contribution to the extragalactic \gr 
background (EGB) can be estimated by writing the EGB intensity as 
$\epsilon I^{sf}_\epsilon \cong R_{\rm H} \zeta \rho {b} L_\gamma/4\pi$, 
where the Hubble radius $R_{\rm H} \cong 4200$ Mpc for a Hubble constant of $71$ km s$^{-1}$ Mpc$^{-1}$, 
$\zeta \sim 3$ -- 10 is a cosmological factor accounting for more active
star formation at redshift $z \gtrsim 1$, and $\rho = \rho_3/(1000$ Mpc$^3$) 
is the local space density of normal and star-forming galaxies. The factor $b \cong 0.4$ 
corrects for the intensity at 100 MeV given the $> 100$ MeV luminosity. Writing 
$L_\gamma = 10^{40} L_{40}$ erg s$^{-1}$ gives 
$\epsilon I^{sf}_\epsilon \cong 3.5\times 10^{-10} b \zeta \rho_3 L_{40}$ 
erg cm$^{-2}$ s$^{-1}$ sr$^{-1}$. For $L_*$ galaxies like the Milky Way, $\rho_3 \cong 3$ -- 10, 
and for starburst galaxies like M82 and NGC 253, $\rho_3$ is an order of magnitude smaller (e.g., Scoville 1992). 
At 100 MeV, a diffuse intensity of $\epsilon I^{EGB}_\epsilon(100$ MeV) $\cong 2.4\times 10^{-9}
$ erg cm$^{-2}$ s$^{-1}$ sr$^{-1}$ was measured with EGRET (Sreekumar
et al. 1998), similar to the {\it Fermi} value at 100 MeV (Abdo et al. 2009b).   
Inserting values for $L_\gamma$ from Table~\ref{gal-comp}, one finds that star-forming and starburst galaxies 
could make a significant, $\gtrsim 10$\% contribution to the EGB at 100 MeV, as previously
suggested (Pavlidou \& Fields 2002; Thompson et al. 2007).

Observations with the {\it Fermi} Gamma-ray Space Telescope provide
evidence that GeV emission has been detected from the starburst
galaxy M82, and weaker though still significant evidence for detection
of NGC 253. The {\it Fermi} LAT detections of these galaxies at GeV
energies, together with the recent discovery of $> 700$ GeV $\gamma$ rays from
M82 with VERITAS (Acciari et
al. 2009) and $> 220$ GeV $\gamma$ rays from NGC 253 with H.E.S.S. (Acero et
al. 2009), introduce a new class of \gr sources to \gr astronomy.
Unlike \gr emitting blazars and radio
galaxies powered by supermassive black holes, the evidence presented 
here supports a cosmic-ray origin for 
\gr production in starburst galaxies. Fermi observations over
the upcoming years will improve our knowledge of spectra, variability
properties, and number of \gr bright starburst galaxies, which will also
constitute important targets for observations with planned large Cherenkov telescope observatories 
CTA and AGIS\footnote{CTA - Cherenkov Telescope Array (http:// www.cta-observatory.org); 
AGIS - Advanced Gamma-Ray Imaging System (http://www.agis-observatory.org)}. 

\acknowledgments

The {\it Fermi} LAT Collaboration acknowledges generous ongoing support
from a number of agencies and institutes that have supported both the
development and the operation of the LAT as well as scientific data analysis.
These include the National Aeronautics and Space Administration and the 
Department of Energy in the United States, the Commissariat \`a l'Energie Atomique
and the Centre National de la Recherche Scientifique / Institut National de Physique
Nucl\'eaire et de Physique des Particules in France, the Agenzia Spaziale Italiana
and the Istituto Nazionale di Fisica Nucleare in Italy, the Ministry of Education,
Culture, Sports, Science and Technology (MEXT), High Energy Accelerator Research
Organization (KEK) and Japan Aerospace Exploration Agency (JAXA) in Japan, and
the K.~A.~Wallenberg Foundation, the Swedish Research Council and the
Swedish National Space Board in Sweden.

Additional support for science analysis during the operations phase is gratefully
acknowledged from the Istituto Nazionale di Astrofisica in Italy and the Centre National d'\'Etudes Spatiales in France.

\clearpage

\begin{table}
\begin{center}
\caption{Results  of  maximum likelihood analyses ({\it gtlike}) of M82
  and NGC 253. }
\label{fit-results} 
\begin{tabular}{ccccccc}
\hline
 & RA$^a$ & Dec$^a$ & $r_{95}^a$ & F($>100$ MeV)$^b$ & photon index$^b$
 & significance$^c$ \\ 
   & (deg)  & (deg) & (deg)& ($10^{-8}$ ph cm$^{-2}$ s$^{-1}$) &  \\
\hline
\hline
M82 & 149.06 & 69.64 & 0.11 & 1.6$\pm 0.5_{\rm stat}\pm 0.3_{\rm sys}$ & 2.2$\pm 0.2_{\rm stat}\pm 0.05_{\rm sys}$ & 6.8 \\
NGC 253 & 11.79 & -25.21 & 0.14 & 0.6$\pm 0.4_{\rm stat}\pm 0.4_{\rm sys}$ & 1.95$\pm 0.4_{\rm stat}\pm 0.05_{\rm sys}$ & 4.8 \\
\hline
\end{tabular}
\end{center}
$^a$Source localization results (J2000) with $r_{95}$ corresponding to the 95\% confidence error radius around the best-fit position.\par 
$^b$Parameters of power-law spectral models fitted to the data:
integrated photon flux $> 100$ MeV and photon index. \par  
$^c$Detection significance of each source.  
\end{table}

\begin{table}\scriptsize
\begin{center}
\caption{Properties of \gr galaxies lacking active central
  nuclei. }
\label{gal-comp}
\begin{tabular}{lllllll}
\hline
 Galaxy & $d$ &	R$_{SN}$ &	$M_{Gas}$ &	${F_{\gamma}}^{a}$ &
 $4{\pi}d^{2}{F_{\gamma}}^{a}$ & ${L_{\gamma}}^{a}$ \\
 & (Mpc) & (yr$^{-1}$) & (10$^9$ M$_\odot$) &	
(10$^{-8}$ ph cm$^{-2}$ s$^{-1}$) & (10$^{42}$ ph s$^{-1}$)  &  (10$^{39}$ erg
 s$^{-1}$) \\
\hline
\hline
LMC$^b$	      & 0.049 $\pm$ 0.001 & 0.005 $\pm$ 0.002 & 0.67 $\pm$ 0.08 & 26.3 $\pm$ 4.7 & 0.074 $\pm$ 0.013 & 0.041 $\pm$ 0.007 \\
Milky Way$^c$ &	1	          & 0.02 $\pm$ 0.01   & 6.5 $\pm$ 2.0	& 4.6 $\pm$ 2.3 & 5.5 $\pm$ 2.8 & 3.2 $\pm$ 1.6\\
M82	      & 3.6 $\pm$ 0.3     & 0.2 $\pm$ 0.1     &	2.5 $\pm$ 0.7 &	1.6 $\pm$ 0.5 & 25 $\pm$ 9 & 13 $\pm$ 5.0\\
NGC 253	      & 3.9 $\pm$ 0.4	  & 0.2 $\pm$ 0.1     &	2.5 $\pm$ 0.6 & 0.6 $\pm$ 0.4 & 11 $\pm$ 7 & 7.2 $\pm$ 4.7\\
\hline
\end{tabular}
\end{center}
$^a$\gr fluxes, $F_{\gamma}$, and luminosities,
$L_{\gamma}$, computed in the
energy range 100 MeV to 5 GeV.\par
$^b$LMC: distance measurement by Pietrzynski et
al. 2009; supernova rate estimated by Tammann, et al.\ 1994; mass
estimate by Bruns et al.\ 2005 (see also Westerlund 1997); \gr
flux from Abdo et al. 2009a.\par
$^c$Gas mass estimate from Dame 1992; \gr flux from
 the Milky Way as viewed from a distance of 1 Mpc; \gr luminosity estimated using
 models which take into account pion-decay, inverse Compton, and
 bremsstrahlung photons produced in both the Galactic disk and halo (Bloemen et
 al. 1984, Strong et al. 2000, Pavlidou \& Fields 2002). 
\end{table}

\end{document}